\documentclass[twocolumn]{aastex6}

\begin{document}
\title{A Search For X-ray Emission From Colliding Magnetospheres In Young Eccentric Stellar Binaries}

\author{Konstantin V.\ Getman and Patrick S.\ Broos}
\affil{Department of Astronomy \& Astrophysics, 525 Davey Laboratory, Pennsylvania State University, University Park PA 16802}
\author{\'Agnes K{\'o}sp{\'a}l}
\affil{Konkoly Observatory, Research Center for Astronomy and Earth Sciences, Hungarian Academy of Sciences, PO Box 67, 1525 Budapest, Hungary}
\author{Demerese M. Salter}
\affil{Department of Astronomy and Laboratory for Millimeter-Wave Astronomy, University of Maryland, College Park, MD 20742, USA}
\and
\author{Gordon P.\ Garmire}
\affil{Huntingdon Institute for X-ray Astronomy, LLC, 10677 Franks Road, Huntingdon, PA 16652, USA}

\begin{abstract}
Among young binary stars whose magnetospheres are expected to collide, only two systems have been observed near periastron in the X-ray band: the low-mass DQ Tau and the older and more massive HD 152404. Both exhibit elevated levels of X-ray emission at periastron. Our goal is to determine whether colliding magnetospheres in young high-eccentricity binaries commonly produce elevated average levels of X-ray activity. This work is based on {\it Chandra} snapshots of multiple periastron and non-periastron passages in four nearby young eccentric binaries (Parenago 523, RX J1622.7-2325 Nw, UZ Tau E, and HD 152404). We find that for the merged sample of all 4 binaries the current X-ray data show an increasing average X-ray flux near periastron (at a $\sim 2.5$-sigma level). Further comparison of these data with the X-ray properties of hundreds of young stars in the Orion Nebula Cluster, produced by the Chandra Orion Ultradeep Project (COUP), indicates that the X-ray emission from the merged sample of our binaries can not be explained within the framework of the  COUP-like X-ray activity. However, due to the inhomogeneities of the merged binary sample and the relatively low statistical significance of the detected flux increase, these findings are regarded as tentative only. More data are needed to prove that the flux increase is real and is related to the processes of colliding magnetospheres.
\end{abstract}

\keywords{stars: individual (Parenago 523, RX J1622.7-2325 Nw, UZ Tau E, HD 152404); stars:pre-main sequence; stars: flare; X-rays: binaries; X-rays: stars}

\section{Introduction} \label{intro_section}

High-amplitude X-ray variability and hard spectra associated with magnetic reconnection flaring are ubiquitous in young stars \citep[e.g.,][]{Feigelson99,Stelzer15}. These frequently ``observed'' X-ray flares exhibit a wide range of durations, from about an hour for faint flares to over a day for bright flares \citep[e.g.,][]{Caramazza07,Getman08a}. The energy distribution of these flares is a power-law with the spectral index similar to that of the Sun \citep{Caramazza07,AlbaceteColombo07,Stelzer07}. This suggests the presence of numerous nano-flares that could be responsible for coronal heating in young stars. Due to insufficient signal such flares remain undetected in the light-curves of young stars making up the ``quiescent'' (a.k.a., ``characteristic'') level of the observed X-ray emission.

The 13-day continuous observation obtained for the {\it Chandra} Orion Ultradeep Project \citep[COUP;][]{Getman05} provided a unique opportunity to study the infrequent large X-ray flares ($<$1~flare/week/star) characteristic of pre-main sequence (PMS) stars \citep{Favata05}. The 200 largest COUP flares rank among the most powerful, longest, and hottest stellar flares known; their inferred coronal structures are the largest reported, comparable to several stellar radii in both disk-bearing (Class II) and diskless (Class III) systems \citep{Getman08a,Getman08b}. The ionization induced in circumstellar disks by these powerful and hard X-rays is expected to significantly influence their chemistry and turbulence (via magneto-rotational instability), perhaps with profound effects on accretion, dust settling, protoplanet migration and other physical processes \citep[e.g.,][]{Ilgner06,Ercolano13}.

The origin of the large COUP flares is unclear. For instance, \citet{Favata05} suggest magnetic loops linking the stellar photosphere with the inner rim of the circumstellar disk. \citet{Getman08a,Getman08b} propose that the majority of the large COUP flares can be viewed as enhanced analogs of the rare solar long-decay events (LDEs) --- eruptive events that produce X-ray emitting arches and streamers with altitudes rising to $>10^5$~km. Both these scenarios involve only a single star.

However, the eccentric short-period PMS binary DQ~Tau has recently been found to exhibit large mm-band and X-ray flares coincident with DQ~Tau's ${\sim}$10~R$_{\star}$ periastron passage \citep{Salter2008,Salter10,Getman11}. These have properties similar to those of large COUP flares, but have been attributed to collisions between the magnetospheres of the binary components. This interpretation is supported by the recurrence of mm flaring in 4 monitored periastron encounters and consistency with a synchrotron model, by the time relationship between the mm and X-ray flares (a Neupert-like effect), and by consistency between the inferred flare loop size and the binary separation. This discovery suggests that some of the large COUP flares could also be produced by colliding magnetospheres. 

Although star forming regions (SFRs) are expected to contain a large fraction of multiple systems \citep[e.g.,][]{Duchene07}, the frequency of close-separation short-period ($P<100$~days) binaries is not well known. At least 10--15\% of young stars in the nearby SFRs Taurus-Auriga, Scorpius-Ophiuchus, and Corona Australis have been found to be such binaries, via spectroscopy \citep{Prato07}. However, only $\sim 60$ PMS binaries with estimated orbital elements have been reported in total (Figure \ref{fig_e_vs_p}), largely because the precise radial velocity measurements of young stars are often difficult to obtain owing to the random and/or systematic contributions by chromospheric and/or accretion activity. Little is known about spectroscopic binaries (SBs) in the Orion Nebula Cluster (ONC). The three known SBs within the COUP field of view (Parenago~1540, 1771, 1925) have incomplete or highly uncertain orbital parameters. 

Among young binary stars whose magnetospheres are expected to collide, only the low-mass DQ~Tau system \citep{Getman11} and the older and more massive HD 152404 system \citep{GomezDeCastro13} have been observed near periastron in the X-ray band. The goal of the current work is to determine whether colliding magnetospheres in young high-eccentricity binaries commonly produce elevated levels of X-ray activity. This work is based on short (3 ks) {\it Chandra} observations of multiple periastron passages in four nearby young eccentric binaries, complemented by observations of the systems away from periastron. In some cases, we monitored the periastrons at optical or near-infrared wavelengths from the ground as well. Our binary sample includes three K-M-type systems (Parenago 523, RX J1622.7-2325, and UZ Tau E) and one older F-type system (HD 152404).

It is important to stress here our realistic expectations about the outcome of this experiment. Within the 60~ks {\it Chandra} X-ray observation of the part of the DQ Tau's periastron, \citet{Getman11} detect one large flare event (with the duration typical to that of the COUP large flares, $\sim 50$~ks) and one much smaller event, superimposed on the large one. Getman et al. also speculate that the large DQ Tau event, in turn, may be a combination of multiple weaker events. Based on these observational results for DQ Tau, we therefore assume that a powerful magnetosphere collision process might release a significant amount of energy stored in large-scale magnetic structures that might result in a flaring activity comprising a wide range of flares. At the same time, as mentioned above, young stars are highly magnetically active objects exhibiting regular ``normal'' X-ray activity powered by magnetic reconnection events on single stars regardless the presence or absence of a binary component. {\it For a young eccentric binary with a strong magnetosphere collision, we thus expect to have a superposition of these two types of flaring activities at periastron.}

Unlike in the \citet{Getman11} study, in the current project our individual 3~ks {\it Chandra} exposures are generally too short for a detection and characterisation of even faint X-ray flares; recall that the typical durations of faint short ``observed'' $Chandra$ X-ray flares are roughly an hour or so \citep[][]{Wolk05,AlbaceteColombo07}. Even if such a $Chandra$ exposure captures the part of the rise/decay phase of a large X-ray flare, it would be impossible to infer the main properties of the parental flare (such as morphology, duration, and energetics). {\it Here, we thus are not so much interested in detailed variability analyses of any ``observed'' flares but rather in comparison of the average levels of X-ray activities between periastron and non-periastron passages.} Further, it should be noted that for individual periastron passages, an elevated level of X-ray activity might not be detected at least for the following two reasons: (astrophysical) a significant flaring activity due to magnetosphere collisions might not take place for every periastron passage, for instance due to the lack of time for the stellar magnetospheres to restore the energies between the passages and/or due to the unfavorable orientation and topologies of the magnetospheres \citep{Salter10}; (observational) if the activity due to a magnetosphere collision is dominated by large flares, but our $3$~ks exposures "land" on inter-flare levels. For the above reasons, our project is set up to sample multiple periastron passages. If a strong flare activity due to magnetosphere collisions happens, we expect to find, on average, a higher X-ray flux near periastrons compared to that of non-periastrons.

The target sample is reviewed in \S \ref{sec_sample} and the {\it Chandra} and the ground-based observations are described in \S \ref{observation_section}. The inferred X-ray photometric and spectral properties are given in \S\S \ref{sec_phot}, \ref{sec_spec}. Comparison of the X-ray binary data with the X-ray data of the ONC PMS stars are presented in \S \ref{sec_sim}. Optical and near-IR results are given in \S \ref{sec_onir_results}. We end with discussion of our new observational findings and directions of further research (\S \ref{sec_discussion}).

\section{OUR SAMPLE OF YOUNG HIGHLY ECCENTRIC BINARIES}\label{sec_sample}

\begin{figure}
\centering
\includegraphics[angle=0.,width=78mm]{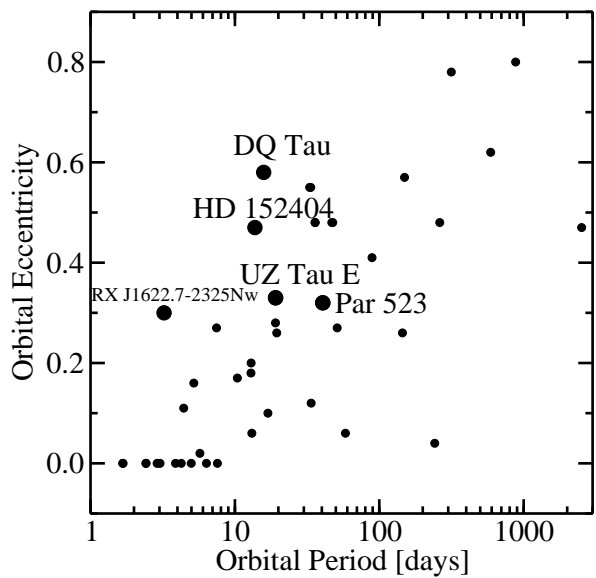}
\caption{Eccentricity versus period for all PMS binaries reported by \citet{Melo01,Prato02,Rosero11}. DQ Tau and our 4 binaries of interest are shown with large points and labeled. \label{fig_e_vs_p}}
\end{figure}

Since the colliding magnetospheres emission mechanism is likely to be sensitive to orbital geometry, we chose to observe systems that are similar to DQ~Tau, specifically those with $e > 0.3$ and $P < 50$~days (Figure \ref{fig_e_vs_p}).  Assuming that magnetospheric interactions require periastron separations and magnetic field strengths comparable to those inferred for the COUP sample of large flares \citep[$B \sim 0.05-0.3$~kG in the outer loop regions, assuming a dipolar topology,][]{Getman08a,Getman08b}, we require targets to have well-established periastron separations $<$15~R$_{\star}$. This criterion excludes the systems RX~J0532.1-0732, 162814-2427, and Par~1925; in Figure \ref{fig_e_vs_p}, these are represented by the three small points next to DQ Tau. Thus, from several objects positioned near DQ Tau on the eccentricity versus period diagram, we are left with the following 4 systems that compose our target sample: Parenago~523, RX~J1622.7-2325 Nw, UZ~Tau~E, and HD 152404.

{\bf Parenago~523}: Within the Orion cloud complex \citep[$D \sim$\ 414~pc; e.g.,][]{Muench08} the {\em ROSAT} source RX~J0530.7-0434 (= 1RXS~J053043.1-043453 = Parenago~523 = V1878 Ori) has been classified as a WTTS (Weak line T-Tauri star) PMS star with spectral type K2--K3 \citep{Alcala96}. This young equal-mass SB2 has a 40.57~day orbital period, an eccentricity of 0.32 \citep{Covino01},  and an inclination of sin~$i \sim 0.98$ \citep{Marilli07}. The stellar mass ($M_1 = M_2 \sim 2$~M$_{\odot}$), radii ($R_1 = R_2 \sim 3.4$~R$_{\odot}$), and rotation period ($P_{1,rot} = P_{2,rot} = 12.9$~days) have been estimated by \citet{Marilli07}. This rotation period was derived by Marilli et al. from the optical light curve of the system that shows rotational modulation with a peak-to-peak amplitude of $\Delta$V$=0.22$~mag; and this period is roughly 1/3 of the orbital period, suggesting non-synchronous rotation. The component separation at periastron is ${\sim}$15~R$_{\star}$ (${\sim}$30~R$_{\star}$  at apastron). Prior to our {\it Chandra} observations, the system had not been observed with modern X-ray telescopes. A {\it ROSAT} observation \citep{Alcala96} provides an estimate of the ``characteristic'' (presumed to be a superposition of small un-resolved flares) intrinsic luminosity of $L_X \sim 1.7 \times 10^{31}$~erg~s$^{-1}$ over 0.5--8~keV, assuming a typical PMS thermal plasma at ${\sim}$2~keV with an extinction of $A_V \sim 0.3$~mag \citep{Covino01,Marilli07}.

{\bf RX~J1622.7-2325 Nw}: Within the $\rho$~Oph SFR \citep[$D = 120$~pc;][]{Loinard08} lies the hierarchical quadruple system RX~J1622.7-2325 \citep{Prato07}. The western component (Nw), a double-lined SB (=SB2), lies $1\arcsec$ from the eastern component (Ne), an $0.1\arcsec$ visual binary. The two systems have been classified as weak-line T-Tauri stars (WTTS) with spectral types M1 and M3, respectively \citep{Martin98}. The Nw system is an equal-mass binary with a 3.23~day orbital period and an eccentricity of 0.3 \citep{Rosero11}, making it the shortest-period high-eccentricity young binary known. \citet{Prato07} has estimated the stellar masses ($M_1 = M_2 \sim 0.6$~M$_{\odot}$) and radii ( $R_1 = R_2 \sim 1.5$~R$_{\odot}$).   The component separation at periastron is only ${\sim}$5~R$_{\star}$ (${\sim}$10~R$_{\star}$ at apastron). Prior to our {\it Chandra} observations, RX~J1622.7-2325 had not been observed with modern X-ray telescopes. A short {\it ROSAT}-PSPC observation away from periastron provides an estimate of the ``characteristic''  intrinsic luminosity for the unresolved quadruple Nw+Ne of $L_X \sim 5 \times 10^{29}$~erg~s$^{-1}$ over 0.5--8~keV, assuming a thermal plasma at ${\sim}$2~keV and a low extinction of $2-4$~mag. This extinction is estimated from the source's position on the $J-H$ vs. $H-K$ diagram, based on our new $JHK$ photometry (\S \ref{onir_observations_section}).


{\bf UZ~Tau~E}: UZ~Tau is a similar hierarchical quadruple system, found in the Taurus-Auriga SFR \citep[$D = 140$~pc;][]{Loinard05}. The eastern (E) SB2 component \citep{Mathieu96} is a non-equal-mass ($M_1,M_2 = 1,0.3$~M$_{\odot}$, $R_1,R_2 = 1.9,1.5$~R$_{\odot}$) binary with a 19.131~day orbital period, an eccentricity of 0.33, and an inclination of $54^\circ$ \citep{Jensen07}; the primary is classified as spectral type M1.  The component separation at periastron is ${\sim}$12~R$_{\star}$ (${\sim}$23~R$_{\star}$ at apastron). Like DQ Tau, UZ~Tau~E is surrounded by an accreting circumbinary disk \citep[e.g.,][]{Jensen07}. The detection of a double-peaked mm-band variability near periastron \citep{Kospal11} indicates a star-star magnetic interaction similar to that of DQ~Tau \citep{Salter10}. An {\it XMM-Newton} observation of UZ~Tau with the E system at orbital phase $\Phi \sim 0.5$ provides an estimate of the ``characteristic'' intrinsic luminosity for the unresolved quadruple UZ~Tau of $L_X = 7.4 \times 10^{29}$~erg~s$^{-1}$ over 0.5--8~keV \citep[XEST project;][]{Gudel07}. No variability (over 45~ks) was detected.

{\bf HD~152404 $=$ AK Sco}: Being a member of the Upper Centaurus-Lupus (UCL) SFR, HD~152404 is much older \citep[$\sim 16$~Myr,][]{Pecaut2012} than our other binary systems, and of later spectral type. We adopt 103~pc for the distance to HD~152404 based on its {\it Hipparcos} parallactic measurement \citep{vanLeeuwen2007}; this is smaller than the average distance to UCL \citep[140~pc,][]{Preibisch08}. Based on strong H$_{\alpha}$ emission and Li~I absorption lines, HD~152404 was identified as a PMS system by \citet{Herbig72}.  \citet{Andersen89} and \citet{Alencar03} show that this is an SB2 binary composed of two equal-mass F5 components ($M_1 = M_2 \sim 1.4$~M$_{\odot}$, $R_1 = R_2 \sim 1.6$~R$_{\odot}$) with a short orbital period (13.6~days), large eccentricity ($e = 0.47$), and an orbit inclination of $i \sim$ 65--70~degrees. The expected component separation near periastron (apastron) is $\sim 11$~R$_{\star}$ ($\sim 30$~R$_{\star}$). The spectral energy distribution (SED) of HD~152404 is fairly typical for a Class II system, fit by a circumbinary disk of about 0.002--0.02~M$_{\odot}$ \citep{Jensen97,Alencar03}. The source extinction is $A_V \sim 0.5$~mag \citep{Alencar03}. Previous X-ray observations of the system include a short 3~ks $Chandra$-ACIS-I observation \citep{Feigelson03,Stelzer06} made at the orbital phase of $\Phi \sim 0.17$, and three recent $\sim$26~ks $XMM$ observations \citep{GomezDeCastro13} performed at the orbital phases $\Phi \sim 0.99,0.15,0.48$. All these X-ray data suggest an atypically soft (for PMS stars) plasma temperature of $\sim 0.5$~keV and an intrinsic X-ray luminosity of $L_X \sim 10^{29}$~erg~s$^{-1}$. \citet{GomezDeCastro13} find that at periastron X-ray fluxes are enhanced by a factor of $\sim$1.5--2 (see their Table~5).

\section{OBSERVATIONS} \label{observation_section}

\subsection{X-ray Observations and Data Extraction} \label{xray_data_reduction_section}

Employing NASA's {\it Chandra X-ray Observatory} \citep{Weisskopf02} we have conducted 17 short ($\sim$3~ks) X-ray observations of our targets near and away from periastron (Table \ref{tbl_xray_photometry}). These were obtained with the ACIS camera \citep{Garmire03}  using one of its back-side illuminated CCDs, ACIS-S3. To mitigate possible photon pileup effects, the {\it Chandra} observations were performed in a sub-array configuration, which reduces the CCD frame time (0.4 s) to 1/8 of nominal.  One archival observation of HD 152404 on a front-illuminated CCD, ACIS-I3, was also analyzed (Table \ref{tbl_xray_photometry}). 

Data reduction follows procedures similar to those described in detail by \citet{Broos10} and \citet[][Appendix B]{Townsley03}. Briefly, using the tool {\it acis\_process\_events} from the CIAO 4.5 software package, the latest available calibration information (CALDB 4.5.8) on time-dependent gain and a custom bad pixel mask are applied, background event candidates are identified, and the data are corrected for CCD charge transfer inefficiency. Using the {\it acis\_detect\_afterglow} tool, additional afterglow events not detected with the standard Chandra X-ray Center (CXC) pipeline are flagged. The event list is cleaned by ``grade'' (only ASCA grades 0, 2, 3, 4, 6 are accepted), ``status'', ``good-time interval'', and energy filters. The slight point-spread function (PSF) broadening from the CXC software position randomizations is removed.

\begin{figure}
\centering
\includegraphics[angle=0.,width=180mm]{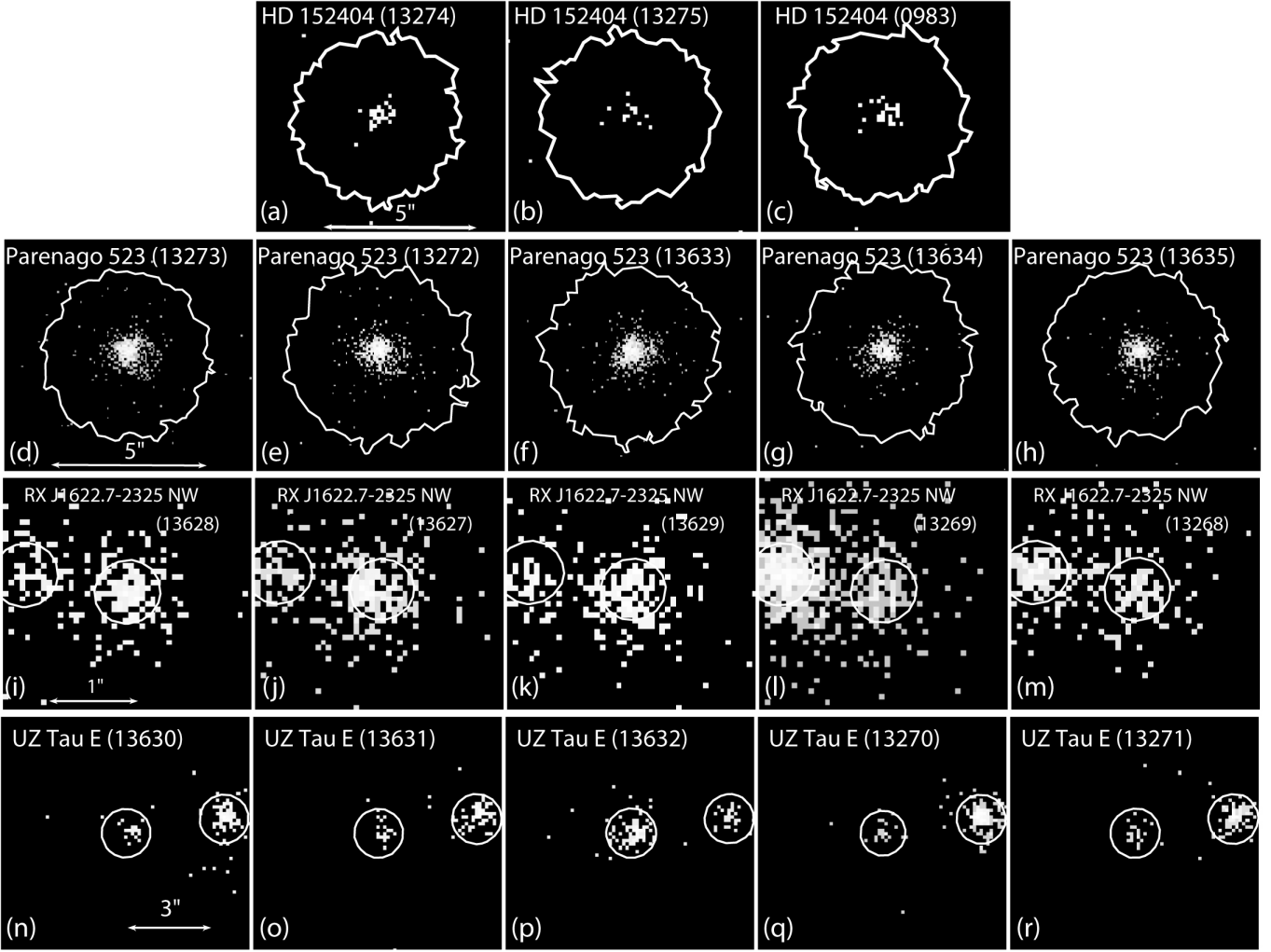}
\caption{For each of the {\it Chandra} observations, X-ray events detected in neighborhood fields around our binaries. The related {\it Chandra} ObdIDs are given in the figure legends. The polygons show the source extraction apertures that enclose 98\%, 98\%, 60\%, and 90\% of the local PSF power for the HD 152404, Parenago 523, RX J1622.7-2325 Nw, and UZ Tau binary systems, respectively. \label{fig_xray_images}}
\end{figure}

Using the ACIS Extract (AE) software package \citep{Broos10,Broos12}, source photons are extracted within polygonal contours enclosing 60--98\% of the local PSF (Figure \ref{fig_xray_images}). In the quadruple system RX J1622.7-2325, {\it Chandra's} unrivaled mirrors easily resolved the Nw and Ne components separated by $1\arcsec$. The background is measured locally in a source-free region using a background algorithm optimized for crowded fields \citep{Broos10}. The AE package was also used to construct source and background spectra, compute redistribution matrix files (RMFs) and auxiliary response files (ARFs), construct light curves and time-energy diagrams, perform a Kolmogorov-Smirnov (K-S) variability test, and compute photometric and spectral properties.

\subsection{Optical and Near-Infrared Photometry} \label{onir_observations_section}   

Optical and near-infrared monitoring of the periastron-passages of our targets were conducted at two observatories with three telescopes: the 1\,m (primary mirror diameter) Ritchey-Chr\'etien-Coud\'e (RCC) telescope and the 60/90/180\,cm (aperture diameter/primary mirror diameter/focal length) Schmidt telescope of the Konkoly Observatory (Hungary), and the 60\,cm (primary mirror diameter) Rapid Eye Mount (REM) telescope at La Silla (Chile). The 1\,m RCC is equipped with a 1300$\times$1340 pixel Roper Scientific WersArray: 1300B CCD camera (pixel scale: 0$\farcs$306), and a Bessel $UBV(RI)_C$ filter set. The Schmidt telescope is equipped with a 4096$\times$4096 pixel Apogee Alta U16 CCD camera (pixel scale: 1$\farcs$03), and a Bessel $BV(RI)_C$ filter set. The REM telescope hosts two parallel instruments: the near-infrared camera REMIR, and the visible camera ROS2. At the time of our observations, only REMIR was operational. REMIR is a Hawaii\,I camera with a useful area of 512$\times$512 pixels, pixel scale of 1$\farcs$2, and $JHK^{\prime}$ filters.

Parenago\,523 was monitored with the Schmidt telescope on two nights, 9/10 and 10/11 October, 2012, and with the 1\,m RCC telescope on two other nights, 19/20 and 20/21 November, 2012. In both cases the observations bracketed the periastron that happened during daytime. UZ\,Tau was monitored with the 1\,m RCC telescope on two nights, 11/12 and 13/14 November 2012 (the periastron happened during daytime for this source as well). RX\,J1622.7--2325 was monitored with the REM telescope during one night, 07/08 July 2012, covering the periastron.

All images were reduced with custom-made IDL scripts. Reduction steps for the CCD images were the usual bias correction, dark subtraction and flatfielding. The near-infrared images were taken in five dither positions, and these were combined to eliminate the sky signal and correct for flatfield differences. For Parenago 524, aperture photometry was obtained for the target in a 6$''$ radius aperture with sky annulus between 10$''$ and 15$''$. The separation of the E and W components of UZ\,Tau is 3$\farcs$8, which means that in some frames the components are well separated, while in others, they are blended together, depending on the actual seeing. We estimated the brightness of the W component with a small, 1$\farcs$0 aperture on the frames with the best seeing, then obtained photometry for the total UZ\,Tau\,E+W system using a large, 12$''$ aperture, and subtracted the contribution of the W component. The separation of RX\,J1622.7--2325\,Nw+Ne and RX\,J1622.7--2325\,S is 13$''$, thus, they are well distinguished in our images. We used an aperture radius of 3$\farcs$6, and sky annulus between 20$\farcs$4 and 24$''$ to obtain photometry for a source that consists of the Nw and Ne components.

Photometric calibration for the optical data was done using the UCAC4 catalog magnitudes \citep{Zacharias2013} of nearby stars in the field (seven stars for Parenago\,523 and two stars for UZ\,Tau), where we first converted the Sloan $r$ and $i$ magnitudes to Johnson-Cousins $R$ and $I$ magnitudes using the formulae of \citet{Jordi2006}. The near-infrared data was calibrated using 2MASS magnitudes \citep{Cutri2003} of two nearby comparison stars.

\section{RESULTS} \label{sec_results}

\subsection{X-ray Photometric Properties and Their Variations with Orbital Phase} \label{sec_phot}

The apparent X-ray photometric flux, $F_X$ (photon cm$^{-2}$ s$^{-1}$), and the median energy of the observed X-ray events, $ME$ (keV), are the principal quantities we use to conduct a homogeneous and systematic science analysis of our four binary systems. Since these systems exhibit low absorption, $F_X$ serves as a good surrogate for intrinsic flux and luminosity.  $ME$ is a known surrogate for both X-ray column density and plasma temperature in PMS stars \citep{Getman10}.   

Observed $F_X$ and $ME$ values are listed in Table \ref{tbl_xray_photometry} for the individual {\it Chandra} observations, for the  ``combined periastron'' and ``combined non-periastron'' observations, as well as for the merged sample of all binary systems. 

\begin{figure}
\centering
\includegraphics[angle=0.,width=150mm]{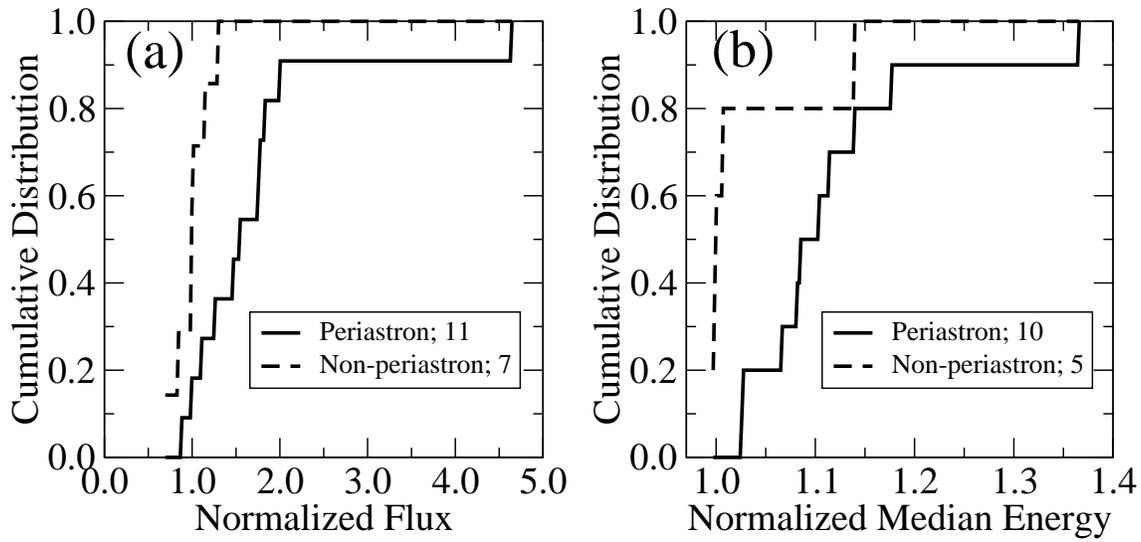}
\caption{For the merged sample of all binaries, the cumulative distributions of the normalized flux (a) and median energy (b) for periastron (solid) and non-periastron (dashed) events samples. \label{fig_flux_me_vs_orbital}}
\end{figure}

{\it The merged sample of all binaries (last 4 rows of the table) is constructed under the assumption that our 4 targets are representative of a class of young eccentric binaries.}  For this merged sample, the $F_X$ and $ME$ data have been normalized by their respective ``combined non-periastron'' averages. For instance, Parenago 523's normalized fluxes are: $82.69/46.54$, $68.13/46.54$, $82.21/46.54$, $52.15/46.54$, and $46.54/46.54$ for ObsIDs 13273, 13272, 13633, 13634, and 13635, respectively. The merged sample comprises 11 periastron and 7 non-periastron $F_X$ measurements (Column 8), and 10 periastron and 5 non-periastron $ME$ measurements (Column 9; omitting HD~152404 \footnote{For the HD 152404 system, its $ME$ values were omitted from the merged periastron and non-periastron samples because HD 152404 was observed with both back-illuminated and front-illuminated CCDs, which have significantly different spectral responses. Unlike $F_X$, $ME$ remains instrumental dependent quantity.}). Figure \ref{fig_flux_me_vs_orbital} presents the cumulative distributions of these normalized fluxes and median energies for the periastron and non-periastron event samples. 

Table~\ref{tbl_xray_photometry} shows that for each of the 4 binary systems, the scatter in $F_X$ and $ME$ across most related periastron or non-periastron X-ray observations are larger than the statistical uncertainties due to Poisson noise ($\sigma_{stat}$). This scatter is likely due to the astrophysical effect of X-ray variability. The uncertainty on sample mean based on the amount of variation around the mean (commonly denoted as the standard error of mean) is defined as the sample standard deviation divided by the square root of the sample size, $[(N-1)^{-1}\sum_{i=1}^{N}(x_i-\hat{x})^2]^{1/2}/(N)^{1/2}$ \citep[equation (4.14) in][]{Bevington1992}. For a sample drawn from a normal distribution the standard error would further indicate the 68\% confidence interval of the mean. Mean estimates and their standard errors are listed for the ``combined periastron'' and ``combined non-periastron'' observations (when $N>1$) as well as for the merged sample of all binary systems. Also provided are the estimates of the medians and their uncertainties derived using the bootstrap technique described in \citet[][their \S3.3]{Getman2014} (last two rows in Table~\ref{tbl_xray_photometry}). 

The following results are evident from Table~\ref{tbl_xray_photometry} and Figure~\ref{fig_flux_me_vs_orbital}. 

\begin{enumerate}

\item For the individual binary systems, the mean value of the X-ray flux near periastron seems higher than that away from periastron. For instance, the ratio of the mean periastron to non-periastron fluxes is 1.5, 1.6, 2.2, and 1.8 for Parenago~523, RX~J1622.7-2325 Nw, UZ Tau E, and HD~152404, respectively.  For all but one system (HD~152404), the mean value of $ME$ near peariastron seems higher than that away from periastron. However, due to the poor observation sampling this result is regarded as a tentative indication.

Specifically, for Parenago~523 and HD~152404, only a single observation is available at the non-periastron and periastron states, respectively. This prohibits estimation of standard errors in these cases.

For RX~J1622.7-2325 Nw, considering the available standard errors on mean $F_X$, the mean X-ray flux near periastron appears 1.6 times higher (with statistical significance $P>98$\%; 2.4-$\sigma$ difference) than that away from periastron. For its mean $ME$ near periastron, the statistical error derived using the error propagation approach \citep[][their \S3.2]{Bevington1992} is consistent with the standard error. However, for its mean $ME$ away from periastron, its statistical error exceeds the standard error.

For UZ Tau E, the statistical errors on the mean $ME$ exceed the standard errors (in cases of both periastron and non-periastron states), as do the errors on the mean flux away from periastron.

\item To reduce flux uncertainties, the 4 systems were merged together. The merged samples comprise 11 periastron and 7 non-periastron normalized (as described above) $F_X$ measurements, and 10 periastron and 5 non-periastron normalized $ME$ measurements. Two analyses of the data suggest that our merged class of young eccentric binaries shows on average brighter and harder X-ray emission near periastron.

First, there appear to be differences in the mean values of $F_X$ and $ME$ (with respect to their standard errors) between the periastron and non-periastron samples at the significance level of over 2-$\sigma$: for the flux the statistical significance is $P>98$\% corresponding to 2.4-$\sigma$ difference, and for $ME$, $P>99$\% corresponding to 2.9-$\sigma$ difference. The same results are obtained if the standard errors on the means are substituted with bootstrap uncertainties \citep[][their \S3.3]{Getman2014}. Similarly, the median values of $F_X$ and $ME$, and their bootstrap uncertainties, show $2.5$-$\sigma$ differences ($P>98$\%), both for the flux and median energy.

Second, the cumulative distributions (CMDs) in Figure \ref{fig_flux_me_vs_orbital} suggest that normalized $F_X$ and $ME$ increase near periastron. The two-sample, two-sided Kolmogorov-Smirnov (K-S) test, which evaluates the null hypothesis that the two samples (periatron and non-periatron in our case) come from the same underlying distribution \citep[e.g.,][]{Wilcox12}, is well-suited to evaluate the strength of that claim.

The K-S test is a two step procedure: first, it determines the statistic $D$ (the maximum value of the absolute difference between two CMDs); second, it calculates the significance level ($p-$value) of the observed value $D$. Many old algorithms for estimating $p-$value given in the literature are approximations that are asymptotically correct as the sample size $N$ approaches infinity but could be unreliable for small $N$ \citep[][and references therein]{Simard2012}. We use the
$ks$\footnote{The $ks$ program is part of the $WRS$ package. The description of the program can be found in \citet{Wilcox12}. The description of the package is available at https://cran.r-project.org/web/packages/WRS2/index.html.} and $ks.boot$\footnote{The description of the program can be found at http://www.inside-r.org/packages/cran/Matching/docs/ks.boot.} programs from the $WRS$ and $Matching$ packages of the R statistical software system that offer the option of computing ``exact'' $p$-values, perfectly correct for any $N$. The recursive \citep{Wilcox12} and bootstrap \citep{Sekhon2011} methods are employed in these calculations. Both programs produce similar results: under the null hypothesis, the $p-$value of the $F_X$ data is $\la 3$\% and the $p-$value for the $ME$ data (excluding HD 152404 as explained above) is $\la 1$\%. These low $p-$values lead us to reject the null hypothesis ($F_X$ and $ME$ are independent of orbital phase).

We also tested the null hypothesis using the Anderson-Darling (AD) statistic. Unlike the K-S test, which is sensitive to the global changes between the CMDs of two datasets, the AD test captures both the global and local differences, i.e., is more sensitive in the distribution tails\footnote{See a related article by Feigelson and Babu at https://asaip.psu.edu/Articles/beware-the-kolmogorov-smirnov-test.}. We use the $ad.test$ program\footnote{The description of the program can be found at http://www.inside-r.org/packages/cran/kSamples/docs/ad.test.} from the $kSamples$ package of the R statistical software system. The program offers the option of computing both ``asymptotic'' \citep{Scholz1987} and ``exact'' \citep{Knuth2011} $p$-values. The ``asymptotic'' method still serves well for sample sizes as low as $N \ga 5$ \citep{Scholz1987}. In agreement with the K-S test, the AD test gives small probabilities that the ``periastron'' and ``non-periastron'' samples are drawn from the same distribution, with $p-$values (``asymptotic'' similar to ``exact'') of $\la 2$\% and $\la 1.5$\% for the $F_X$ and $ME$ data, respectively.

\end{enumerate}

In summary, for each of our 4 binary systems, there is a tentative indication that the mean X-ray flux near periastron is higher than that away from periastron; however, the poor data sampling prohibits the evaluation of the X-ray flux uncertainties. For the merged sample of all 4 binaries, two independent analyses provide an indication (at the significance level of $\sim 2.5$-$\sigma$) that the average level of X-ray emission near periastron is higher and the emission is on average harder than that away from periastron.

\subsection{Simulations Testing Hypothesis of COUP-like X-ray Activity}\label{sec_sim}

In this section we carry out a simulation based on the rich X-ray dataset for PMS stars in ONC to test the null hypothesis that the ``periastron vs. non-periastron'' X-ray flux variations seen in our binary systems could be explained by the ``normal'' PMS X-ray activity observed in numerous stellar members of ONC. Here, we more formally evaluate the significance of our results by defining a statistic that quantifies the apparent periastron flux enhancement: the ratio of the ``combined periastron'' average X-ray flux to ``combined non-periastron'' average X-ray flux ($F_{p\_comb}/F_{np\_comb}$; hereafter referred to as $F_p/F_{np}$).  We then evaluate the statistical significance of the observed value of $F_p/F_{np}$ for each of the binaries and for the merged sample of all 4 binaries\footnote{Notice that the statistics to evaluate the apparent periastron flux enhancement is not restricted to the use of the $F_p/F_{np}$ ratio quantity; the $F_p - F_{np}$ difference quantity, normalized to the distance squared, could equally well be used instead.}.  

In this statistical hypothesis test, our null hypothesis is that all our observations arise from normal X-ray variability in single PMS stars.  We simulate our observed flux ratios ($F_p/F_{np}$, Column 2 in Table \ref{tbl_simulation_predictions}) under the null hypothesis via Monte Carlo sampling of the light curves observed for hundreds of PMS stars in the Orion Nebula Cluster (ONC), produced by the Chandra Orion Ultradeep Project \citep[COUP;][]{Getman05}.

From the ``lightly-obscured'' sample of COUP PMS stars \citep{Feigelson05}, most of which are ONC members, we select the following mass-stratified sub-samples of ONC PMS stars with known stellar mass estimates from \citet{Getman05}: 447 stars with $M<3$~M$_{\odot}$, 60 stars with $1<M<3$~M$_{\odot}$, 387 stars with $M<1$~M$_{\odot}$, 53 stars with $0.5<M<0.7$~M$_{\odot}$, 314 stars with $M<0.5$~M$_{\odot}$, 191 stars with $0.2<M<0.4$~M$_{\odot}$. Notice that formally the mass strata $1<M<3$~M$_{\odot}$, $1<M<3$~M$_{\odot}$, and $0.5<M<0.7$~M$_{\odot}$ could be viewed as most appropriate mass sub-samples to compare with the X-ray data of the equal-mass binaries Parenago 523, HD 152404, and RX J1622.7-2335 NW, respectively. However, for the following reasons, we prefer to present the simulation results that include all 6 sets of mass-strata for every binary: firstly, UZ Tau is a highly non-equal mass binary; secondly, the $1<M<3$~M$_{\odot}$ COUP mass sub-sample composed of mainly young ($\la$ few Myr) G-F-type stars could be non-characteristic of the much older ($\sim 16$~Myr) F-type stars in HD 152404; thirdly, it is simply interesting to test the dependence of the simulation results on the choice of a mass sub-sample.

\begin{figure}
\centering
\includegraphics[angle=0.,width=180mm]{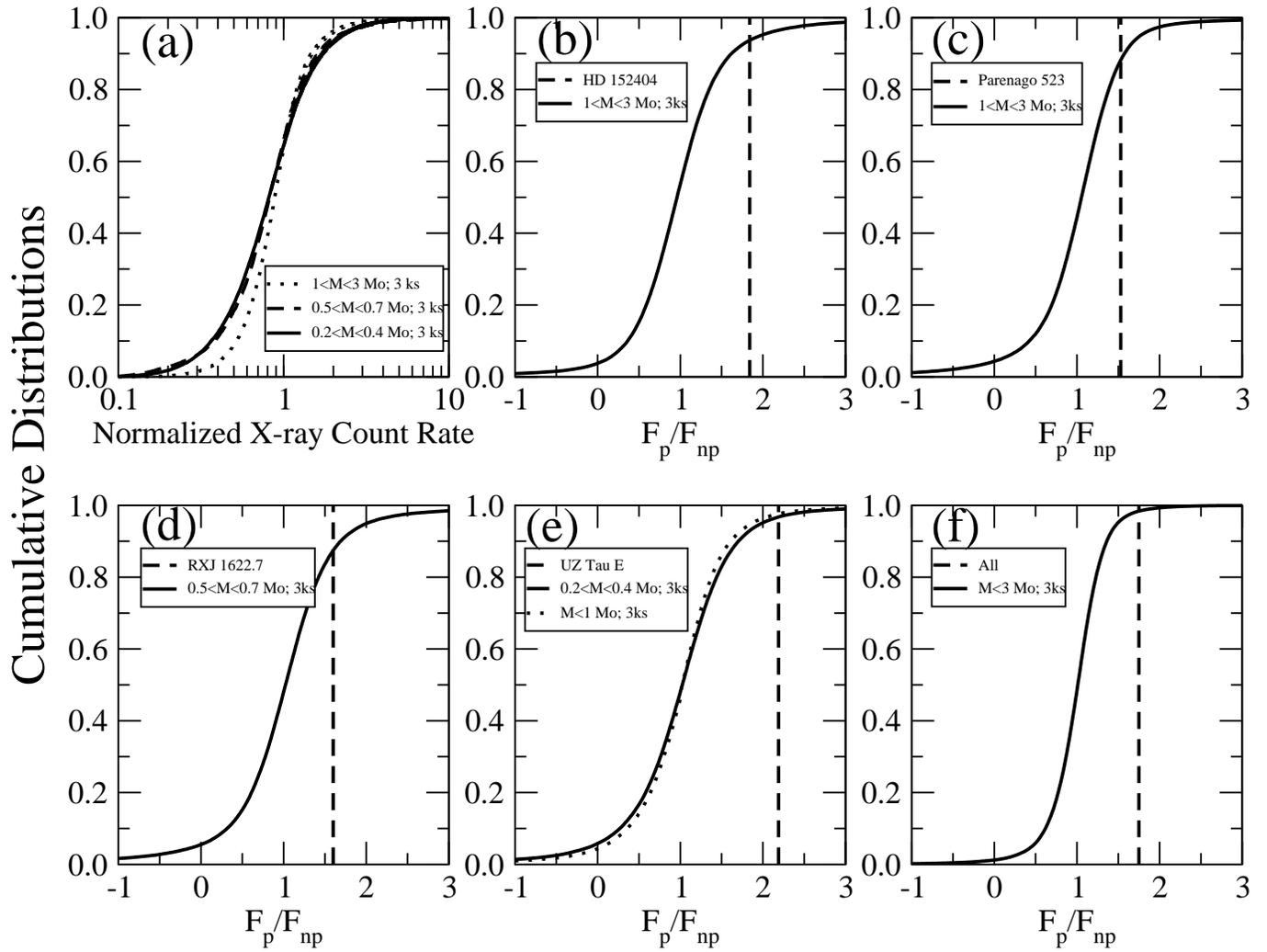}
\caption{Results of the COUP simulations. (a) The cumulative distributions of normalized (as described in the text) X-ray count rate for the three mass-stratified COUP PMS samples. Panels (b-f) exemplify the inferred probabilities for detecting the flux ratios $F_p/F_{np}$ (Column 2 in Table~\ref{tbl_simulation_predictions}), using mass-stratified COUP PMS samples within the mass ranges characteristic of our binary systems. \label{fig_simulation_results}}
\end{figure}

Conceptually, each COUP mass sub-sample simulation consists of two stages. In the first stage we construct a cumulative distribution function (CDF) for the ``normalized'' count rate (NCR) produced by the stars in the sub-sample. That is, for each COUP source within the sub-sample, its photon arrival time-series is divided into numerous chunks of duration $dt$. Note that the duration of the source chunks selected for the simulations ($dt$) ranged from 3 to 50~ks, to account for sensitivity differences among the COUP and our binary observations, due to the differences in distance. The count rates measured in each of these chunks are normalized by the mean count rate of the source. The distribution of those ``normalized'' measurements from all the chunks in all the sources within the sub-sample represents the sub-sample's CDF. For instance, the CDFs inferred for the three COUP sub-samples ``$1<M<3$'', ``$0.5<M<0.7$'', and ``$0.2<M<0.4$'' are shown in Figure \ref{fig_simulation_results}a. Some differences in the NCR are present between the $<0.7$~M$_{\odot}$ and $>1$~M$_{\odot}$ ONC stars. For instance, $NCR < 0.4$ is seen in 16\% and $7$\% of the $M<0.7$~M$_{\odot}$ and $M>1$~M$_{\odot}$ stars, respectively; $NCR > 2$ is seen in 6\% and 3\% of the $M<0.7$~M$_{\odot}$ and $M>1$~M$_{\odot}$ stars, respectively. 

In the second stage, for each of our four binaries, we constructed synthetic $F_p/F_{np}$ quantities assuming the null hypothesis that their X-ray emission arises from ``COUP-PMS like X-ray activity unrelated to binarity''. We use $N_p$ and $N_{np}$ quantities as inputs to the simulations describing the observing program used for each binary (Table \ref{tbl_simulation_predictions}).  Here, $N_p$ and $N_{np}$ are numbers of available ``periastron'' and ``non-periastron'' X-ray datasets. This is done in the following way: from the pool of the COUP NCR distributions (derived earlier, separately for the 6 COUP mass sub-samples and the 6 trial versions of $dt$; Table \ref{tbl_simulation_predictions}), we randomly draw $100000$ samples of $N_p$ and $N_{np}$ NCR sequences to construct COUP CDFs for the random variable $F_p/F_{np}$. The significance of each of our observed $F_p/F_{np}$ statistics --- characterised by the probability of obtaining an $F_p/F_{np}$ value more extreme (greater) than observed, under the null hypothesis --- can be directly obtained from the appropriate simulated CDFs shown in Figure \ref{fig_simulation_results}.

The derived probabilities for each of the binaries, as well as for the combined sample of all 4 binaries, are given in Table \ref{tbl_simulation_predictions}. For UZ Tau E and the combined sample of ``All 4 systems'', the inferred probabilities are less than 5\%, while for HD~152404, Parenago~523, and RX J1622.7-2325 NW, the probabilities are greater than 5\%. Within the framework of the statistical hypothesis test, these results imply that the null hypothesis of COUP PMS-like X-ray activity can be confidently rejected for UZ Tau E and the combined sample of ``All 4 systems''. Expressed in another way, when the binaries are considered separately from each other, there is a $\sim 10-20$\% chance of random coincidence between the ``periastron passage'' and the observed ``higher average level of X-ray emission'' for the three out of four binaries, consistent with ``normal X-ray activity'' unrelated to binarity. However, for the merged sample of ``eccentric young binaries'', the effect of the higher average level of X-ray activity near periastron becomes too significant to be explained by the ``normal X-ray activity'' model and requires an additional component of X-ray activity near periastron.

Finally, the simulations were modified and repeated assuming that for $50$\% of the COUP PMS stars their observed X-ray emission represents a combination of X-rays from unresolved equal-mass binary systems. In this case, the modified null hypothesis is that the X-ray activity in our 4 binary systems of interest is similar to ``COUP-PMS like X-ray activity that is unaffected by periastron passages in binaries'' (i.e., assuming that most of the COUP binaries have nearly circular orbits). The principal result of the new simulations remains similar to the original ones: the null hypothesis can be confidently rejected for UZ Tau E and the combined sample of ``All 4 systems''.

\subsection{X-ray Spectroscopy}\label{sec_spec}

The source and background spectra for the merged periastron and non-periastron X-ray data along with the related calibration files were produced by AE \citep{Broos10}. Using the {\it XSPEC} package (Arnaud 1996), these spectra were fitted with two-temperature (Parenago 523) or one-temperature (other systems) APEC plasma emission models (Smith et al. 2001). Coronal elemental abundances characteristic of young stars \citep[0.3 times solar;][]{Getman05} were assumed. X-ray absorption was modeled using the WABS model of atomic cross sections of \citet{Morrison83}. The fits were performed using the $\chi^{2}$ statistic.

\begin{figure}
\centering
\includegraphics[angle=0.,width=80mm]{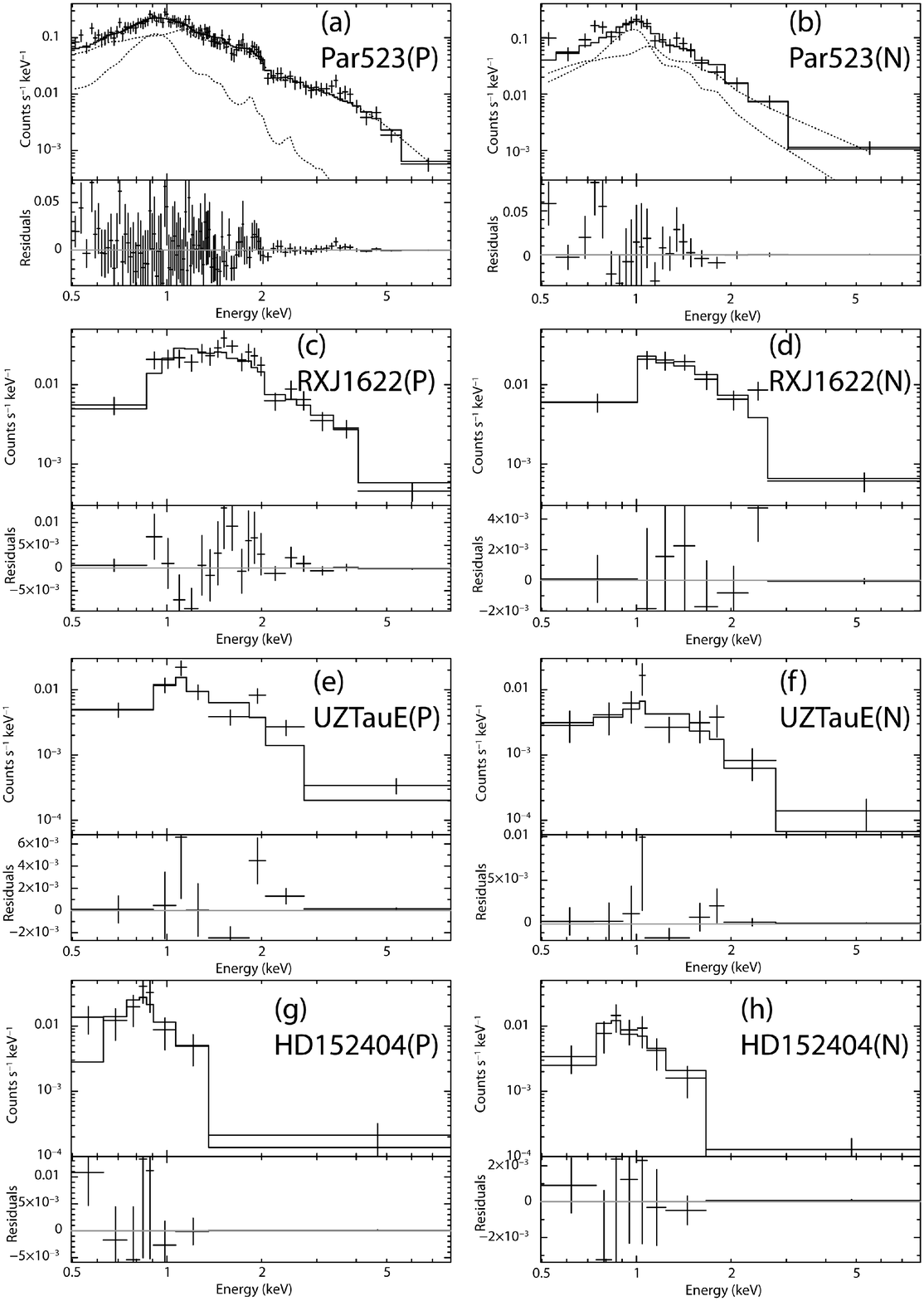}
\caption{Fits of the {\it Chandra} spectra: Parenago 523 (a,b), RX J1622.7-2325 Nw (c,d), UZ Tau E (e,f), and HD 152404 (g,h). Spectra of all available combined periastron and non-periastron observations are shown on the left and right panels, respectively. The two-temperature (Parenago 523) and one-temperature (the remaining sources) model fits (lines) to the data (crosses) are presented. For UZ Tau E and HD 152404, the model fits `Model1' and `Model3' (Table \ref{tbl_spectra}) are shown, respectively. \label{fig_spec}}
\end{figure}

X-ray spectral fitting of {\it Chandra} imaging data for young stars is commonly ambiguous. Qualitatively different spectral models might fit the data reasonably well. For instance, \citet{Getman05} describe the ambiguity in fitting the exceptionally deep COUP data for hundreds of Orion young stars. Here in Table \ref{tbl_spectra} and Figure \ref{fig_spec}, for Parenago 523 and RX J1622.7-2325 Nw, we choose to present results for the `best model' families that resemble those found for the majority of the COUP PMS stars, that is $(kT_1,kT_2) \sim$ (0.7--1,2--3)~keV and $kT \sim$ 2--5~keV for 2-T and 1-T COUP models, respectively. In fact, by varying initial parameters and executing the {\it error} and {\it steppar} commands, we were unable to find an alternative `best fit' model family for RX J1622.7-2325 Nw. Meanwhile, there is a wide range of numerous alternative models that `successfully' fit the poor data of UZ Tau E and HD 152404. To exemplify this ambiguity, in Table \ref{tbl_spectra} we choose to present a few `best fit' alternative models with physically reasonable values of plasma temperature and column density. The main findings of this section are:


\begin{enumerate}

\item{\bf Parenago 523 and RX J1622.7-2325 Nw.} In line with the photometry results (\S \ref{sec_phot}), the spectral fitting indicates higher plasma temperatures and elevated (by a factor of $[1.7-1.9]$) intrinsic X-ray luminosities ($L_X$) near periastron ({\it although these variations are not statistically significant}). Our estimates of X-ray luminosity outside periastron for the individual components of these equal-mass binaries ($L_X = 7 \times 10^{30}$~erg s$^{-1}$ for a $2$~M$_{\odot}$ component in Par 523; $L_X = 4 \times 10^{29}$ ~erg s$^{-1}$ for a $0.6$~M$_{\odot}$ component in RX J1622.7-2325Nw) are fully consistent with the location of Taurus and ONC PMS stars of similar masses on the X-ray luminosity - mass diagram \citep{Telleschi2007, Preibisch2005}. 


\item{\bf UZ Tau E.} In line with the photometry results --- constant $ME$ but elevated $F_X$ near periastron --- the various alternative model fits presented in Table \ref{tbl_spectra} indicate small variations in the plasma temperature and/or column density, and $\ga 2$ times elevated levels of intrinsic X-ray luminosity near periastron ({\it although not statistically significant}). The derived here estimate for the total X-ray luminosity for this non-equal mass binary (M$_1 \sim 1$~M$_{\odot}$, M$_2 \sim 0.3$~M$_{\odot}$) outside periastron ($L_X \sim 10^{29}$ erg s$^{-1}$) is at the lower boundary of the $0.3-1$~M$_{\odot}$ PMS locus on the  X-ray luminosity vs. mass diagram \citep{Telleschi2007, Preibisch2005}. This is consistent with the effect of suppression of hard X-ray emission in many accreting versus non-accreting PMS stars.

\item{\bf HD 152404.} In agreement with the {\it XMM} results \citep{GomezDeCastro13}, our analysis of the {\it Chandra} spectrum gives a moderate X-ray luminosity ($L_X \sim 10^{29}$~erg s$^{-1}$) and a soft plasma temperature ($\sim 0.6$~keV) for this  F5$+$F5-type binary system. It is important to emphasize here that these X-ray characteristics are consistent with those of F-type MS stars \citep[e.g.,][]{Panzera1999}. The relatively old age of HD 152404 ($\sim 16$~Myr) and the agreement of its X-ray properties with those of old F-type stars support the notion that the bulk of X-rays observed from HD 152404 might be produced through processes related to magnetic activity.


\end{enumerate}


\subsection{Optical and Near-Infrared Results}\label{sec_onir_results}
 
\begin{figure}
\centering
\includegraphics[angle=0.,width=180mm]{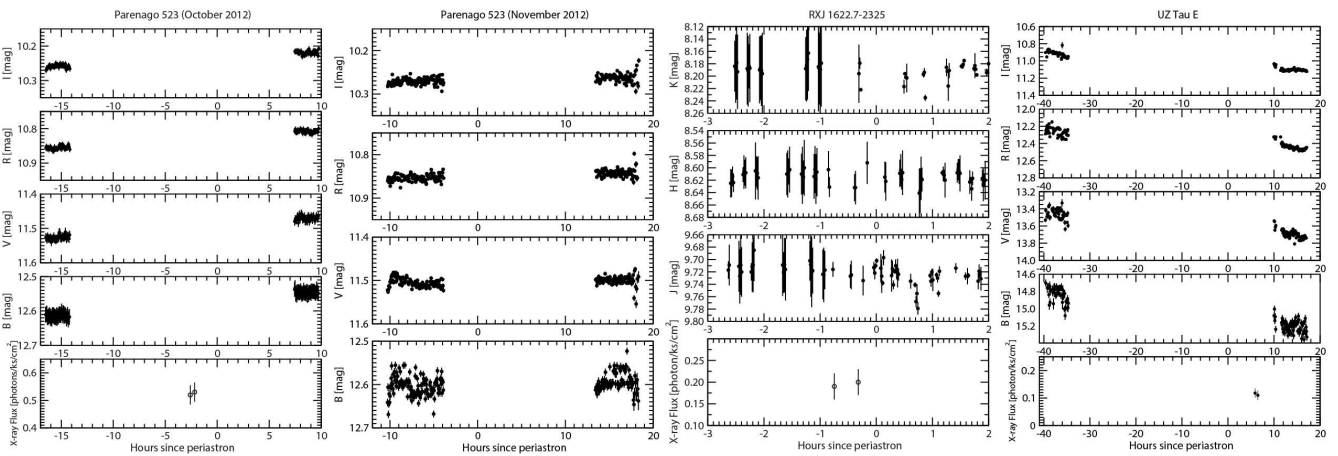}
\caption{Optical, NIR, and X-ray lightcurves for (from left to right, respectively): Parenago 523, RX J1622.7-2325 (this quadruple system remains unresolved in NIR), and UZ Tau E. The lower panels show the X-ray lightcurves for the X-ray observations that overlap or are nearly contiguous with the ONIR data: ObsId 13634 for Parenago 523, ObsID 13629 for RX J1622.7-2325, and ObsId 13632 for UZ Tau E. \label{fig_onir}}
\end{figure}

\citet{Salter10} found evidence for systematic optical brightenings of DQ\,Tau almost coincident with the periastrons. The brigtenings are apparent in all $V$, $R$, and $I$ filters, with the largest amplitude in $V$. They speculate that the brightenings may be connected to the accretion process: accretion pulses synchronized with the binary orbit are predicted by model simulations \citep{Artymowitz1996,Shi2012}, and observed as well (e.g., UZ Tau E; \citealt{Jensen07}). Alternatively, the optical brightenings may be due to the magnetospheric reconnection events. Motivated by this, we monitored three of our targets from the ground at optical or near-infrared wavelengths around their periastrons. The resulting light curves are plotted in Figure \ref{fig_onir}.

{\bf Parenago 523} was observed twice, around two different periastrons. Its $BVRI$ photometry does not show large variability, the peak-to-peak amplitude is about 0.1\,mag in all four filters. This is less than what \citet{Marilli07} found, probably because we sampled only a small part of the total 12.9\,d period of the binary. Thus, we conclude that the observed brightness variations are consistent with rotational modulation of a spotted stellar surface.

{\bf RX\,J1622.7--2325} was monitored in the $JHK$ filters. This source is a weak-line T\,Tauri, with no evidence for infrared excess at these wavelengths due to disk emission. Therefore, the data reflect the brightness of the stellar photospheres. The object was mostly constant within the measurement uncertainties, except for a slight dip about $0.7-0.9$ hours after periastron, best visible in the $J$-band data (the different filters were not simultaneously used). We speculate that this may be an eclipsing event, with the caveat that it is a quadruple system not resolved by our photometric observations, which complicates the interpretation of the results. In eccentric binaries, there is a higher probability to have eclipses close to periastron. RX\,J1622.7--2325 has not been known as an eclipsing binary before. These data may be the first hint that it shows eclipses. The dip we detected is only $\Delta{}J$=0.06\,mag deep, detected to about 6$\sigma$, thus, follow-up observations are needed to confirm the eclipsing binary nature of the system.

{\bf UZ\,Tau\,E} is well known as a variable star at optical and infrared wavelengths (e.g., \citep{Xiao2012, Kospal2012}). \citet{Kospal11} reported optical monitoring of UZ\,Tau\,E at periastron and reported peak-to-peak variability of 1.19, 0.96, and 0.57 mag in the $V$, $R$, and $I$ band, respectively, and given the similarity of the optical light curves to the millimeter-wavelength light curve, they speculate that apart from variable accretion rate, strong magnetic activity may also contribute to the optical flux changes. As Figure \ref{fig_onir} shows, we again detected optical variability in UZ\,Tau\,E, with peak-to-peak amplitudes of 0.8, 0.5, 0.4, and 0.3 mag in $B$, $V$, $R$, and $I$, respectively. The colors of the variability in the present study are very similar to those in \citet{Kospal11}, suggesting similar physical mechanisms.

In no cases did we see any sign for a brightening close to periastron similar to that observed in DQ\,Tau by \citet{Salter10}.

\section{Conclusions}\label{sec_discussion}

Two major mechanisms for optical/radio/X-ray activity in young high eccentric binaries have been discussed in the literature. The first mechanism, magnetic activity due to colliding magnetospheres, was considered to explain  the optical/mm/X-ray activity in DQ Tau \citep{Salter10,Getman11}, optical/mm in UZ Tau E \citep{Kospal11}, and radio activity in V773 Tau A \citep{Massi08,Adams11}. As a manifestation of the magnetic activity due to colliding magnetospheres the X-ray flux near periastron of a young high eccentric binary is expected to be higher and generally harder than that away from periastron \citep{Getman11}. The second major mechanism,  pulsed accretion from binary-disk interactions, was invoked to explain the optical/NIR activity in DQ Tau \citep{Mathieu97,Bary14} and optical in UZ Tau E \citep{Jensen07}. See for instance \citet{Bary14} and references therein, for details regarding different pulsed  accretion scenarios.

Our goal is to determine whether colliding magnetospheres in young high-eccentricity binaries commonly produce elevated average levels of X-ray activity. The current work is based on {\it Chandra} snapshots of multiple periastron and non-periastron passages in four nearby young eccentric binaries (Parenago 523, RX J1622.7-2325 Nw, UZ Tau E, and HD 152404). For three of these binaries, X-ray emission was observed (Parenago 523, RX J1622.7-2325 Nw) or resolved (UZ Tau E) here with a modern X-ray telescope for the first time; their average intrinsic X-ray properties are presented in \S \ref{sec_spec}.

We believe that the current X-ray dataset is too sparse to clearly reveal the effects of colliding magnetospheres in our binaries, but two findings presented here certainly encourage additional observations.  X-ray photometric analysis (\S \ref{sec_phot}) shows that for the merged binary sample X-ray flux near periastron is higher and the emission is harder (at the significance level of $\sim 2.5$-$\sigma$) than that away from periastron. The COUP simulations (\S \ref{sec_sim}) show that for the merged binary sample, as well as the individual system UZ Tau E, the X-ray flux variations between the ``periastron'' and ``non-periastron'' states can not be explained by the ``normal'' X-ray activity observed in numerous PMS members of ONC.

However, the following opposing arguments may cast some doubts on the supporting ideas above. First, at this relatively low $\sim 2.5$-$\sigma$ level the result of the increased and hardened X-ray emission near periastron could still be spurious. Second, our sample of binaries was chosen to include systems with a certain orbital geometry, such as high eccentricity, relatively low orbital period, and periastron separations limited to $15$~R$_{\star}$ (\S \ref{sec_sample}). The latter was chosen to be comparable to the coronal loop sizes inferred for some large COUP flares \citep{Getman08a} to insure that interacting magnetospheres at such proximities would be capable of producing strong X-ray flares. Nevertheless, the range of the periastron separations among our binaries is wide, [$5-15$]~R$_{\star}$. Other parameters, such as stellar mass, accretion rate, and age are also known to affect the X-ray emission of PMS stars  \citep[][and references therein]{Getman2014}. All these and other parameters, many of which drastically vary across our binary sample, might affect the production of X-rays (both ``normal'' X-rays and X-rays from colliding magnetospheres) in different ways. This reasoning might cast doubt on the validity of combining the 4 binaries in a single merged sample. Although, with regards to one of the most influential parameters on X-ray emission (stellar mass), our COUP simulations show that the results remain similar among different mass strata (Table \ref{tbl_simulation_predictions}).

Future coordinated multi-wavelength observation campaigns of young eccentric binaries, spanning both a larger number of periastron passages and a larger range of orbital phase, are thus desirable to search for the excess X-ray emission and provide better understanding of the underlying mechanisms. With the assumption that the X-ray flux increase found in \S \ref{sec_phot} is real and is persistent at a similar level for each of the individual binaries, we suggest that as many as $>50$ ``periastron'' and $>50$ ``non-periastron'' short ($3$~ks) passages of individual binary systems to be observed in X-rays in order to derive meaningful information (at $5$-$\sigma$ significance level) on the presence/absence of the excess X-ray emission. The disk-free systems Parenago~523 and RX~J1622.7-2325 Nw  could be considered as good observation candidates since these cases allow disentangling from the effects of accretion. Parenago~523 system is also advantageous for its extreme brightness in X-rays, allowing good statistics in each of the short X-ray observations (Table~\ref{tbl_xray_photometry}). Meanwhile, RX~J1622.7-2325 Nw is preferred for its relatively short orbital period ($3.23$~days) to allow a swift observation campaign. In addition, a potential unique feature of RX~J1622.7-2325 Nw is that due to the very small component separation, even at apastron ($\sim 10$~R$_{\odot}$), the system might experience strong magnetosphere interactions throughout the entire orbital phase. The disky system UZ Tau E could be considered as a valuable observation candidate as well since the results of our COUP simulations for this system indicate inconsistency of its X-ray emission with the model of ``normal'' X-ray activity.

\acknowledgements

We thank Leisa Townsley (Penn State University) for her role and financial support in the development of the wide variety of ACIS data reduction and analysis techniques and tools used in this study. We thank Eric Feigelson (Penn State University), Scott Gregory (St Andrews University), Gaitee Hussain (ESO), and Costanza Argiroffi (University of Palermo) for helpful discussions. We thank the CXC staff for their help with the {\it Chandra} observation scheduling. We also thank the anonymous referees for their time and useful comments that helped to improve this work. This work is supported by the {\it Chandra} GO grant SAO GO2-13016X (K. Getman, PI) and  the {\it Chandra} ACIS Team contract SV4-74018 (G.~Garmire \& L.~Townsley, Principal Investigators), issued by the {\it Chandra} X-ray Center, which is operated by the Smithsonian Astrophysical Observatory for and on behalf of NASA under contract NAS8-03060. The Guaranteed Time Observations (GTO) included here were selected by the ACIS Instrument Principal Investigator, Gordon P. Garmire, of the Huntingdon Institute for X-ray Astronomy, LLC, which is under contract to the Smithsonian Astrophysical Observatory; Contract SV2-82024. \'A.K. acknowledges the support from the Momentum grant of the MTA CSFK Lend{\"u}let Disk Research Group and the Hungarian Research Fund OTKA grant K101393.\\

\vspace{5mm}
\facilities{CXO, Konkoly(RCC), LaSilla(REM)}

\software{CIAO (\url{http://cxc.harvard.edu/ciao/}), CALDB (\url{http://cxc.harvard.edu/caldb/}), ACIS Extract \citep{Broos10,Broos12}, IDL, ks (\url{https://cran.r-project.org/web/packages/WRS2/index.html}), ks.boot (\url{https://cran.r-project.org/web/packages/Matching/Matching.pdf}), ad.test (\url{https://cran.r-project.org/web/packages/kSamples/kSamples.pdf}), XSPEC \citep{Arnaud96}}

\clearpage
\newpage

\floattable 
\begin{deluxetable}{cccccccccc}
\centering \tabletypesize{\scriptsize} \rotate \tablewidth{0pt} \tablecolumns{10}

\tablecaption{X-ray Photometry \label{tbl_xray_photometry}}

\tablehead{\colhead{Object} &  \colhead{ObsID} & \colhead{Start Time} & \colhead{$\Phi$} & \colhead{$NC$, $\sigma_{stat}$} & \colhead{PSF} &\colhead{Expo.} & \colhead{$F_X$, $\sigma_{stat}$} & \colhead{$ME$, $\sigma_{stat}$} & \colhead{$P_{NoVar}$}\\
\colhead{} &  \colhead{} & \colhead{UT} & \colhead{} & \colhead{(counts)} &\colhead{} & \colhead{(sec)} & \colhead{($10^{-5}$ phot cm$^{-2}$ s$^{-1}$)} & \colhead{(keV)} & \colhead{\%}\\
\colhead{(1)} & \colhead{(2)} & \colhead{(3)} & \colhead{(4)} & \colhead{(5)} & \colhead{(6)}  & \colhead{(7)}  & \colhead{(8)} & \colhead{(9)} & \colhead{(10)}}

\startdata
Parenago 523 & 13273 & 2011-12-31T22:32:15 & $(1.0056-1.0064)$ & $692.3,26.8$ & 0.98 & 2728 & $82.69,3.21$ & $1.25,0.02$ & 75\\
Parenago 523 & 13272 & 2012-01-01T17:03:20 & $(1.0247-1.0256)$ & $571.5,24.4$ & 0.98 & 2729 & $68.13,2.91$ & $1.13,0.02$ & 38\\
Parenago 523 & 13633 & 2012-02-10T00:30:02 & $(0.9936-0.9945)$ & $701.5,27.0$ & 0.98 & 2783 & $82.21,3.17$ & $1.15,0.02$ & 25\\
Parenago 523 & 13634 & 2012-10-10T15:00:53 & $(0.9975-0.9984)$ & $444.6,21.6$ & 0.98 & 2784 & $52.15,2.54$ & $1.09,0.03$ & 69\\
Combined   & ... & ...   & Periastron &  ...        & ... & ... & $71.29\pm 7.22$ & $1.15\pm0.03$ & ...\\
Parenago 523 & 13635 & 2012-12-04T15:51:31 & $(0.3538-0.3547)$ & $396.1,20.4$ & 0.98 & 2783 & $46.54,2.40$ & $1.06,0.02$ & 25\\
Combined   & ... & ...   & Non periastron &  ...        & ... & ... & $46.54\pm$... & $1.06\pm$... & ...\\
\cline{1-10}
RX J1622.7-2325 Nw & 13628 & 2012-04-07T14:15:10 & $(0.9819-0.9929)$ & $113.1,11.2$ & 0.60 & 2785 & $23.67,2.34$ & $1.66,0.06$ & 76\\
RX J1622.7-2325 Nw & 13627 & 2012-06-11T03:49:31 & $(0.9702-0.9809)$ & $141.7,12.5$ & 0.58 & 2784 & $30.44,2.68$ & $1.63,0.06$ & 97\\
RX J1622.7-2325 Nw & 13629 & 2012-07-07T01:40:14 & $(0.9900-1.0009)$ & $ 91.9,10.2$ & 0.59 & 2784 & $19.39,2.14$ & $1.54,0.09$ & 99\\
Combined   & ... & ...   & Periastron &  ...        & ... & ... & $ 24.50\pm 3.22$ & $1.61\pm0.04$ & ...\\
RX J1622.7-2325 Nw & 13269 & 2012-04-25T06:43:04 & $(0.4546-0.4654)$ & $ 83.5,10.7$ & 0.60 & 2784 & $17.47,2.23$ & $1.50,0.07$ & 1.3\\
RX J1622.7-2325 Nw & 13268 & 2012-07-09T01:28:33 & $(0.6067-0.6176)$ & $ 62.3,8.5$ & 0.60 & 2785 & $13.03,1.78$ & $1.51,0.14$ & 57\\
Combined   & ... & ...   & Non periastron &  ...        & ... & ... & $ 15.25\pm 2.22$ & $1.50\pm0.005$ & ...\\
\cline{1-10}
UZ Tau E & 13630 & 2012-01-12T03:50:11 & $(0.9895-0.9911)$ & $ 18.9,4.9$ & 0.91 & 2729 & $ 2.51,0.65$ & $1.29,0.45$ & 8.2\\
UZ Tau E & 13631 & 2012-08-09T21:39:24 & $(1.0056-1.0072)$ & $ 16.9,4.6$ & 0.90 & 2784 & $ 2.22,0.61$ & $1.58,0.35$ & 85\\
UZ Tau E & 13632 & 2012-11-13T16:57:24 & $(1.0134-1.0151)$ & $ 89.9,10.0$ & 0.91 & 2785 & $11.70,1.30$ & $1.32,0.09$ & 64\\
Combined   & ... & ...   & Periastron &  ...        & ... & ... & $ 5.48\pm 3.11$ & $1.40\pm0.09$ & ...\\
UZ Tau E & 13270 & 2011-12-27T18:43:48 & $(0.1861-0.1876)$ & $ 18.8,4.9$ & 0.89 & 2729 & $ 2.54,0.66$ & $1.16,0.24$ & 1.5\\
UZ Tau E & 13271 & 2012-01-07T09:04:24 & $(0.7402-0.7416)$ & $ 18.8,4.9$ & 0.91 & 2729 & $ 2.50,0.65$ & $1.32,0.25$ & 2.6\\
Combined   & ... & ...   & Non periastron &  ...        & ... & ... & $ 2.52\pm 0.02$ & $1.24\pm0.08$ & ...\\
\cline{1-10}
HD 152404 & 13274 & 2012-07-17T19:51:35 & $(0.9986-1.0008)$ & $ 33.8,6.3$ & 0.98 & 2785 & $ 3.97,0.75$ & $0.85,0.04$ & 2.4\\
Combined   & ... & ...   & Periastron             &  ...        & ... & ... & $ 3.97\pm$... & $0.85\pm$... & ...\\
HD 152404 & 13275 & 2012-06-24T07:21:00 & $(0.2699-0.2723)$ & $ 12.8,4.1$ & 0.98 & 2783 & $ 1.50,0.48$ & $1.10,0.15$ & 57\\
HD 152404 & 0983 & 2001-08-19T14:56:38 & $(0.1727-0.1749)$ & $ 23.9,5.4$ & 0.98 & 3109 & $ 2.82,0.64$ & $0.90,0.05$ & 27\\
Combined   & ... & ...   & Non periastron             &  ...        & ... & ... & $ 2.16\pm 0.66$ & footnote 1 & ...\\
\cline{1-10}
Object & ... & ... & $\Phi$ & ... & ... & ... & mean($F_X/F_{X,nonPer})$ & mean($ME/ME_{nonPer}$)& ...\\
\cline{1-10}
All systems merged & ... & ... & Periastron & ... & ... & ... & $1.75\pm0.31$ & $1.10\pm0.02$& ...\\
All systems merged & ... & ... & Non Periastron & ... & ... & ... & $1.00\pm0.07$ & $1.00\pm0.02$& ...\\
\cline{1-10}
Object & ... & ... & $\Phi$ & ... & ... & ... & median($F_X/F_{X,nonPer})$ & median($ME/ME_{nonPer}$)& ...\\
\cline{1-10}
All systems merged & ... & ... & Periastron & ... & ... & ... & $1.55\pm0.21^{a}$ & $1.07\pm0.02^{a}$& ...\\
All systems merged & ... & ... & Non Periastron & ... & ... & ... & $1.00\pm0.08^{a}$ & $1.00\pm0.02^{a}$& ...\\
\cline{1-10} 
\enddata

\tablecomments{X-ray Photometry. Column 1: Name of a binary system. Column 2: ObsID values of the X-ray observations from the $Chandra$ Observation Catalog. Column 3: Start times of the X-ray observations in UT. Column 4: Range of the orbital phase covered by the X-ray observation. Columns 5-7: X-ray photometry results, including X-ray net counts ($NC$) within source extraction region, fraction of the point spread function (PSF) enclosed by the source extraction region, and effective X-ray exposure in seconds. The photometric quantity $NC$ and its 68\% confidence interval are given for the $(0.5-8)$~keV energy band.  Columns 8-9: X-ray photometry results, including apparent X-ray flux ($F_X$) in photons cm$^{-2}$ s$^{-1}$ and X-ray median energy ($ME$) in keV. These quantities are given for the $(0.5-8)$~keV energy band. For the individual observations, the 68\% confidence intervals reported for $F_X$ and $ME$ incorporate the statistical component (Poisson noise) omitting any additional systematic components (such as variability). For the ``combined'' periastron and non-periastron states, the reported uncertainty is a standard error on mean, calculated as the ratio of the sample standard deviation to the square root of the sample size (see the text). Column 10: K-S probability statistic under the no-variability null hypothesis within a single X-ray observation (i.e., small values suggest variability). Note (a): the last two rows of this table list the medians of the combined fluxes and energies, normalized to the average values of their respective non-periastron states. The reported uncertainties on these medians were derived using the bootstrap apporach given in \citet{Getman2014}.}
\end{deluxetable}

\floattable 
\begin{deluxetable}{ccccccccccc}
\centering \tabletypesize{\scriptsize} \rotate \tablewidth{0pt} \tablecolumns{11}

\tablecaption{COUP Simulation Predictions \label{tbl_simulation_predictions}}

\tablehead{\colhead{Object} &  \colhead{$F_p/F_{np}$} & \colhead{$N_p$} & \colhead{$N_{np}$} & COUP & Pb$_3$ & Pb$_{10}$ & Pb$_{20}$ & Pb$_{30}$ & Pb$_{40}$ & Pb$_{50}$\\
\colhead{} &  \colhead{} & \colhead{} & \colhead{} & \colhead{Sample} &\colhead{(\%)} & \colhead{(\%)} & \colhead{(\%)} & \colhead{(\%)} & \colhead{(\%)} & \colhead{(\%)}\\
\colhead{(1)} & \colhead{(2)} & \colhead{(3)} & \colhead{(4)} & \colhead{(5)} & \colhead{(6)}  & \colhead{(7)}  & \colhead{(8)} & \colhead{(9)} & \colhead{(10)} & \colhead{(11)}}

\startdata
Parenago 523 & 1.53 &  4 &  1 & $M < 3$~M$_{\odot}$ &  13.8 &  17.1 &  16.1 &  15.2 &  14.7 &  14.1\\
Parenago 523 & 1.53 &  4 &  1 & $1 < M< 3$~M$_{\odot}$ &  12.0 &  10.9 &  10.2 &   9.4 &   9.3 &   8.6\\
Parenago 523 & 1.53 &  4 &  1 & $M < 1$~M$_{\odot}$ &  14.7 &  18.0 &  17.2 &  16.5 &  15.4 &  14.8\\
Parenago 523 & 1.53 &  4 &  1 & $0.5 < M < 0.7$~M$_{\odot}$ &  19.1 &  17.7 &  17.1 &  16.4 &  15.6 &  15.3\\
Parenago 523 & 1.53 &  4 &  1 & $M < 0.5$~M$_{\odot}$ &  18.9 &  18.4 &  17.4 &  16.6 &  15.6 &  15.0\\
Parenago 523 & 1.53 &  4 &  1 & $0.2 < M < 0.4$~M$_{\odot}$ &  20.4 &  18.3 &  17.6 &  16.8 &  15.7 &  15.2\\
RX J1622.7-2325 NW & 1.60 &  3 &  2 & $M < 3$~M$_{\odot}$ &   9.7 &  11.0 &  10.4 &   9.9 &   9.1 &   8.8\\
RX J1622.7-2325 NW & 1.60 &  3 &  2 & $1 < M< 3$~M$_{\odot}$ &   7.5 &   6.8 &   6.5 &   5.6 &   5.8 &   5.3\\
RX J1622.7-2325 NW & 1.60 &  3 &  2 & $M < 1$~M$_{\odot}$ &  10.2 &  11.7 &  11.4 &  10.9 &  10.1 &   9.5\\
RX J1622.7-2325 NW & 1.60 &  3 &  2 & $0.5 < M < 0.7$~M$_{\odot}$ &  12.7 &  11.8 &  11.5 &  10.9 &  10.3 &  10.1\\
RX J1622.7-2325 NW & 1.60 &  3 &  2 & $M < 0.5$~M$_{\odot}$ &  12.9 &  12.1 &  11.3 &  10.6 &  10.2 &   9.5\\
RX J1622.7-2325 NW & 1.60 &  3 &  2 & $0.2 < M < 0.4$~M$_{\odot}$ &  13.4 &  12.1 &  11.5 &  10.8 &  10.2 &   9.8\\
UZ Tau E & 2.19 &  3 &  2 & $M < 3$~M$_{\odot}$ &   2.2 &   2.8 &   2.7 &   2.3 &   2.1 &   1.9\\
UZ Tau E & 2.19 &  3 &  2 & $1 < M< 3$~M$_{\odot}$ &   1.8 &   1.5 &   1.5 &   1.1 &   1.4 &   1.2\\
UZ Tau E & 2.19 &  3 &  2 & $M < 1$~M$_{\odot}$ &   2.4 &   3.1 &   2.9 &   2.6 &   2.3 &   2.0\\
UZ Tau E & 2.19 &  3 &  2 & $0.5 < M < 0.7$~M$_{\odot}$ &   3.7 &   3.6 &   3.1 &   2.9 &   2.7 &   2.4\\
UZ Tau E & 2.19 &  3 &  2 & $M < 0.5$~M$_{\odot}$ &   3.1 &   3.0 &   2.9 &   2.5 &   2.2 &   2.0\\
UZ Tau E & 2.19 &  3 &  2 & $0.2 < M < 0.4$~M$_{\odot}$ &   3.3 &   3.0 &   2.8 &   2.5 &   2.2 &   2.1\\
HD 152404 & 1.84 &  1 &  2 & $M < 3$~M$_{\odot}$ &   8.7 &   8.9 &   8.6 &   8.0 &   7.9 &   7.5\\
HD 152404 & 1.84 &  1 &  2 & $1 < M< 3$~M$_{\odot}$ &   6.4 &   5.7 &   5.6 &   5.3 &   5.0 &   4.6\\
HD 152404 & 1.84 &  1 &  2 & $M < 1$~M$_{\odot}$ &   9.3 &   9.4 &   8.9 &   8.5 &   8.4 &   8.1\\
HD 152404 & 1.84 &  1 &  2 & $0.5 < M < 0.7$~M$_{\odot}$ &   9.7 &   8.8 &   8.6 &   8.5 &   8.2 &   8.2\\
HD 152404 & 1.84 &  1 &  2 & $M < 0.5$~M$_{\odot}$ &  10.5 &   9.7 &   9.0 &   8.6 &   8.5 &   8.0\\
HD 152404 & 1.84 &  1 &  2 & $0.2 < M < 0.4$~M$_{\odot}$ &  10.8 &   9.7 &   9.1 &   8.8 &   8.7 &   8.4\\
All 4 systems & 1.75 & 11 &  7 & $M < 3$~M$_{\odot}$ &   1.6 &   2.5 &   1.9 &   1.7 &   1.3 &   1.0\\
All 4 systems & 1.75 & 11 &  7 & $1 < M< 3$~M$_{\odot}$ &   1.7 &   1.6 &   1.1 &   0.9 &   0.6 &   0.3\\
All 4 systems & 1.75 & 11 &  7 & $M < 1$~M$_{\odot}$ &   1.8 &   2.6 &   2.0 &   1.8 &   1.4 &   1.1\\
All 4 systems & 1.75 & 11 &  7 & $0.5 < M < 0.7$~M$_{\odot}$ &   3.8 &   3.7 &   2.7 &   1.8 &   1.5 &   1.1\\
All 4 systems & 1.75 & 11 &  7 & $M < 0.5$~M$_{\odot}$ &   2.6 &   2.6 &   2.0 &   1.9 &   1.5 &   1.2\\
All 4 systems & 1.75 & 11 &  7 & $0.2 < M < 0.4$~M$_{\odot}$ &   2.8 &   2.5 &   2.0 &   1.8 &   1.4 &   1.2\\
\enddata 
 
\tablecomments{COUP Simulation Predictions. Column 1: Name of a binary system. Column 2: Ratio of ``periastron'' to ``non-periastron'' fluxes, $F_p /F_{np}$; these fluxes are combined observed X-ray fluxes for the ``periastron'' and ``non-periastron'' states, respectively (see Table~1). For instance, for HD~152404, $F_p = 3.97$ and $F_{np} = 2.16$ in flux units (see Table~1), hence, $F_p/F_{np} = 1.84$. For each of the binaries, the COUP simulations produce probability distributions of $F_p/F_{np}$. Columns 3 and 4. Number of available ``periastron'' and ``non-periastron'' X-ray observations (see Table~1). These are considered as parameters in the COUP simulations. Column 5: COUP stellar sample used in the simulations. Columns 6-11: Probabilities for detecting the flux ratio $F_p/F_{np}$ (Column 2) assuming the null hypothesis of the ``COUP-PMS like X-ray activity unrelated to binarity''. The columns give results of the simulations with differently chosen time binning for the COUP count rate time series: Pb$_3$ is a probability from the COUP simulations that use $3$~ks COUP time segments, Pb$_{10}$ - $10$~ks segments,  Pb$_{20}$ - $20$~ks segments, Pb$_{30}$ - $30$~ks segments, Pb$_{40}$ - $40$~ks segments, and Pb$_{50}$ - $50$~ks segments.}
\end{deluxetable}

\floattable 
\begin{deluxetable}{cccccccc}
\centering \tabletypesize{\scriptsize} \rotate \tablewidth{0pt} \tablecolumns{8}

\tablecaption{X-ray Spectroscopy \label{tbl_spectra}}

\tablehead{\colhead{Object} &  \colhead{Orbital} & \colhead{$N_H$} & \colhead{$kT$} & $EM$ & $L_X$ & $\chi^2_{red}$ & dof\\
\colhead{} &  \colhead{State} & \colhead{($10^{22}$~cm$^{-2}$)} & \colhead{(keV)} & \colhead{(10$^{52}$~cm$^{-3}$)} &\colhead{(10$^{31}$~erg~s$^{-1}$)} & \colhead{} & \colhead{}\\
\colhead{(1)} & \colhead{(2)} & \colhead{(3)} & \colhead{(4)} & \colhead{(5)} & \colhead{(6)}  & \colhead{(7)}  & \colhead{(8)}}

\startdata
Parenago~523 & Periastron & $<0.01$ & $0.94^{+0.05}_{-0.06}$;$3.41^{+0.32}_{-0.27}$ & $40.7^{+6.3}_{-7.1 }$;$182.6^{+10.6}_{-9.3 }$ & 2.7 & 0.7 & 99\\
Parenago~523 & Non-Periastron & $<0.01$ & $1.05^{+0.19}_{-0.09}$;$2.43^{+1.05}_{-0.54}$ & $55.6^{+38.5}_{-18.3}$;$82.2^{+22.9}_{-41.2}$ & 1.4 & 1.2 & 18\\
RX J1622.7-2325 Nw & Periastron & $0.44^{+0.11}_{-0.08}$ & $3.13^{+0.60}_{-0.52}$ & $11.9^{+1.9}_{-1.4}$ & 0.14 & 1.1 & 17\\
RX J1622.7-2325 Nw & Non-Periastron & $0.42^{+0.17}_{-0.11}$ & $2.17^{+0.66}_{-0.42}$ & $8.0^{+1.9}_{-1.5}$ & 0.08 & 1.2 & 5\\
UZ Tau E & Periastron; Model1 & 0.14: & 2.0f & 2.4: & 0.024: & 3.0 & 6\\
UZ Tau E & Non-Periastron; Model1 & $<0.01$: & 2.0f & 0.8: & 0.008: & 0.9 & 7\\
UZ Tau E & Periastron; Model2 & 0.09: & 3.0: & 2.2: & 0.025: & 3.5 & 5\\
UZ Tau E & Non-Periastron; Model2 & 0.09f & 3.5: & 1.0: & 0.012: & 0.8 & 7\\
UZ Tau E & Periastron; Model3 & $<0.01$f & 4.3: & 1.8: & 0.024: & 3.2 & 6\\
UZ Tau E & Non-Periastron; Model3 & $<0.01$: & 4.4: & 0.8: & 0.011: & 0.8 & 6\\
HD 152404 & Periastron; Model1 & 0.37: & 0.28: & 8.3: & 0.030: & 1.2 & 5\\
HD 152404 & Non-Periastron; Model1 & 0.23: & 0.67: & 1.2: & 0.007: & 0.5 & 5\\
HD 152404 & Periastron; Model2 & 0.08f & 0.58 & 1.3: & 0.007: & 0.9 & 6\\
HD 152404 & Non-Periastron; Model2 & 0.08f & 0.78: & 0.7: & 0.004: & 0.4 & 6\\ 
HD 152404 & Periastron; Model3 & 0.09: & 0.55f & 1.4 : & 0.007: & 0.9 & 6\\
HD 152404 & Non-Periastron; Model3 & 0.33: & 0.55f & 1.8: & 0.010: & 0.5 & 6\\
\enddata

\tablecomments{X-ray Spectroscopy. Column 1: Name of a binary system. Column 2: Orbital state. Columns 3-6: Inferred spectral properties: X-ray column density, plasma temperature, emission measure, and X-ray luminosity. 1-$\sigma$ confidence limits are given when appropriate. Two-temperature fits were used for the Parenago 523 data; one-temperature fits --- for the rest of the data. The spectral fitting of UZ Tau E and HD 152404 is ambiguous; examples of a few spectral models that formally fit the data are shown. The ``:'' and ``f'' suffixes denote ``uncertain'' and ``fixed'' values, respectively. Columns 7-8: Reduced $\chi^2$ for the overall spectral fit and degrees of freedom.}
\end{deluxetable}

\end{document}